\documentclass[prl,twoside,twocolumn]{revtex4}
\usepackage{graphicx}
\usepackage{amsmath,amssymb,amsfonts,amsthm}
\usepackage{hyperref}
\usepackage{color}

\newcommand{\tder}[2]{\frac{d#1}{d#2}}
\newcommand{\rmax}{ r_\text{max} } 

\newcommand{\ADN}{ADN}

\begin{document}
\title{Relativistic minimization with constraints: A smooth trip to Alpha Centauri}
\author{Riccardo Antonelli\footnote{riccardo.antonelli92@gmail.com}}
\affiliation{Padua, Italy}
\author{Alexander R. Klotz\footnote{aklotz@mit.edu \\ Author list is alphabetical.}}
\affiliation{Department of Chemical Engineering, Massachusetts Institute of Technology}
\begin{abstract} We consider the relativistic generalization of the problem of the ``least uncomfortable'' linear trajectory from point A to point B. The traditional problem minimizes the time-integrated squared acceleration (termed the ``discomfort''), and there is a universal solution for all distances and durations. This universality fails when the maximum speed of the trajectory becomes relativistic, and we consider the more general case of minimizing the squared proper acceleration over a proper time. The least uncomfortable relativistic trajectory has a rapidity that evolves like the motion of a particle in a hyperbolic sine potential, agreeing with the classical solution at low velocities. We consider the special case of a hypothetical trip to Alpha Centauri, and compare the minimal-discomfort trajectory to one with uniform Earth-like acceleration. \end{abstract}
\maketitle

In a recent paper in the American Journal of Physics,  Anderson, Desaix, and Nyqvist \cite{anderson2016least} (ADN) considered the ``least uncomfortable'' trajectory of linear motion from point A to point B, defined as the path covering a distance $X$ in time $T$ for which the integrated squared acceleration (the ``discomfort'') is minimized. ADN derived an elegant solution, showing that the jerk (third derivative of position) is constant on such a trip, and compared their solution to those found by variational approximation methods. The least uncomfortable solution, as presented by ADN, is universal, in that the velocity relative to its maximum as a function of distance or time relative to the total distance or duration is independent of $X$ and $T$. 

While the ADN solution nominally applies to any trip of distance $X$ over time $T$, if $T$ becomes short enough and $X$ large enough, the maximum speed reached during the trip can approach the speed of light. There is in fact an unstated assumption in the ADN solution that $X/T\ll c$. In this letter we generalize the solution and consider the least uncomfortable \textit{relativistic} journey from A to B.

The magnitude of acceleration depends on the reference frame in which it is measured, be it the ``lab'' inertial frame fixed to the start and end points, or the ``ship'' accelerating frame that moves between them. In the lab frame, as the ship asymptotically approaches the speed of light its coordinate acceleration $\frac{d^2 x}{dt^2}$ approaches zero. In the ship frame, however, the \textit{proper} acceleration dictates the inertial forces experienced on-board. To minimize acceleration-induced discomfort, we wish to minimize the total proper acceleration experienced. This can be minimized considering the total duration of the trip in the lab frame, but because observers in the ship frame are experiencing discomfort, we wish to minimize the squared \textit{proper} acceleration integrated over \textit{proper} time.

In their paper, ADN considered a one-dimensional trip over distance $X$ and time $T$, beginning and ending at rest with velocity $v=0$. Their solution minimizes the discomfort integral:

\begin{equation}
F=\int_{0}^{T}{a^{2}dt}
\label{eq:classF}
\end{equation}
subject to the constraint:

\begin{equation}
T=\int_{0}^{X}{\frac{dx}{v}}.
\label{eq:classC}
\end{equation}
The solution, as derived by ADN, is a cubic function in time:

\begin{equation}
\frac{x}{X}=3\left(\frac{t}{T}\right)^{2}-2\left(\frac{t}{T}\right)^{3}
\label{eq:classical}
\end{equation}
This implies that velocity is quadratic in time (with $v_{max}=3/2\,X/T$), that acceleration is linear, and that jerk is constant. 


In the non-relativistic limit, proper time is equivalent to coordinate time, and proper acceleration to coordinate acceleration and the relativistic disagreement between ``lab'' and ``ship'' clocks need not be considered. However, when considering the least uncomfortable relativistic journey, it is desirable to minimize the ``proper discomfort'' as experienced in the frame of the traveller. In the following, units in which $c=1$ are implicitly used. We define the discomfort functional $F$ as the squared proper acceleration integrated over proper time $\tau$ for a total proper duration $\mathcal{T}$.

\begin{equation}
	F = \int_{0}^{\mathcal{T}} a^2 d\tau
	\label{}
\end{equation}
Since the functional to be minimized depends on the second time derivative of position, it is convenient to phrase the question in terms of a minimization over \emph{velocity} profiles $\beta(\tau)$, so that $F$ is first order in derivatives of $\beta$ and the problem can be solved with the Lagrangian formalism. In light of the dimensionality of the problem we choose to consider the rapidity $r(\tau) = \tanh^{-1} \beta(\tau)$ as the dynamical variable; $F$ is rewritten as
\begin{equation}
	F = \int_0^\mathcal{T} \mathcal{L}_0(r,\dot r) d\tau = \int_0^\mathcal{T} \dot r^2 d\tau 
	\label{}
\end{equation}
where we have identified the Lagrangian $\mathcal{L}_0(r,\dot r)$. The dot $\dot{\,}$ denotes a derivative with respect to proper time $\tau$.

We take boundary conditions such that the velocity is zero at $\tau=0$ and $\tau =\mathcal{T}$:

\begin{equation}
	r(\tau = 0) = 0\,,\quad r(\tau = \mathcal{T}) = 0\,,
	\label{}
\end{equation}
and impose that the total distance travelled (in the lab frame) is $X$ as a constraint:

\begin{equation}
	\int_0^X dx = \int_0^T \beta dt = \int_0^\mathcal{T} \beta \gamma d \tau  = \int_0^\mathcal{T} \sinh r d\tau = X
	\label{}
\end{equation}
This formulation presents the question as a standard constrained optimization problem which can be solved \cite{clarke} through the introduction of a Lagrangian multiplier $\lambda$ for the constraint. In practice, we switch to the minimization of the extended functional

\begin{equation}
	F[r(\tau)] - \lambda \int_0^\mathcal{T} \sinh r(\tau) \, d\tau\,,
	\label{}
\end{equation}
where the minus sign has been introduced for later convenience. This implies an extended Lagrangian 

\begin{equation}
	\mathcal{L} = \left( \tder{r}{\tau} \right)^2 - \lambda \sinh r
	\label{lagra}
\end{equation}
It is immediately recognized \eqref{lagra} is the Lagrangian for a particle moving in a potential

\begin{equation}
	U(r) = \frac{\lambda}{2} \sinh r\,,
	\label{}
\end{equation}
after we have identified the particle's position with $r$ and its velocity with $\tder{r}{\tau}$. The resulting Euler-Lagrange equation is

\begin{equation}
	\ddot r = - \frac{\lambda}{2} \cosh r\,.
	\label{eq:ode}
\end{equation}
We can attempt a solution of the equation of motion \eqref{eq:ode} by introducing a special function. The energy of the particle

\begin{equation}
	E = \frac{1}{2} \dot r^2 + \frac{\lambda}{2} \sinh r
	\label{}
\end{equation}
is conserved, and is in particular $2E = \lambda \sinh \rmax $, with $\rmax$ the maximum rapidity reached at $\tau = \mathcal{T}/2$, which means

\begin{equation}
	\dot r = \pm \sqrt{ \lambda \sinh \rmax  - \lambda \sinh r }
	\label{}
\end{equation}

\begin{equation}
	\Rightarrow \pm \frac{dr}{\sqrt{1 - \dfrac{\sinh r }{\sinh \rmax } } } = \sqrt{\lambda \sinh \rmax }d\tau
	\label{separated}
\end{equation}
The two signs of the radical are for $\tau<\mathcal{T}/2$ and $\tau>\mathcal{T}/2$.

We define the following function:

\begin{equation}
	L(r;\rmax) := \int_0^r \frac{ds}{\sqrt{1- \dfrac{\sinh s}{\sinh\rmax} } }
	\label{}
\end{equation}
From the initial conditions, since $L(0, \rmax) = 0$,

\begin{equation}
	\sqrt{\lambda \sinh(\rmax)}\, \tau = L(r ; \rmax) \,.
	\label{solution}
\end{equation}
Equation \eqref{solution} is the expression for the proper time elapsed $\tau$ as a function of the current rapidity, and the two parameters $(\lambda,\rmax)$. This expression only holds with $0 < \tau < \mathcal{T}/2$, until $r=\rmax$ at $\tau = \mathcal{T}/2$; the remaining section is determined by symmetry. In principle, this can be inverted to obtain $r(\tau ; \lambda, \rmax)$, the general solution to the constrained minimization problem. In practice, for any given value of the parameters $\lambda > 0$, $\rmax > 0$, the function $r(\tau;\lambda,\rmax)$ minimizes the functional $F$ for some values $X$, $\mathcal{T}$ of the total distance and proper time. To obtain the solution relative to given $X$, $\mathcal{T}$, this relationship must also be inverted to reconstruct the correspoding parameters $\lambda$, $\rmax$.

For what concerns $\mathcal{T}$ as a function of $(\lambda,\rmax)$, the total proper time $\mathcal{T}$ can be found as the solution to $r(\mathcal{T}/2) = \rmax$; in accordance with \eqref{solution}:

\begin{multline}
	\mathcal{T} = \frac{2}{\sqrt{\lambda \sinh \rmax  } } L(\rmax ; \rmax ) \\
	= \frac{2}{\sqrt{\lambda \sinh\rmax } } \int_0^{\rmax} \frac{ds}{\sqrt{1-\sinh s /\sinh \rmax  } }\,.
	\label{eq:tintegral}
\end{multline}
Thus $\mathcal{T}$ can be expressed numerically as a function of $(\lambda,\rmax)$ in one integral.

A similar computation yields $X$ as an integral too:

\begin{equation}
	X = \int_0^\mathcal{T} \sinh r  d\tau = 2 \int_0^{\mathcal{T}/2} \sinh r d\tau 
\end{equation}
and then using \eqref{separated}:

\begin{equation}
X	= \frac{2}{\sqrt{\lambda \sinh \rmax  } } \int_0^{\rmax} \frac{\sinh s \, ds}{\sqrt{1- \sinh s / \sinh \rmax }} 
	\label{eq:xintegral}
\end{equation}
Let us verify the results of ADN are recovered in the classical limit. If $\beta \ll 1$, proper acceleration and proper time are equivalent to acceleration and time, $r \sim \beta$, and $\gamma \sim 1$. Thus the Lagrangian \eqref{lagra} reduces to

\begin{equation}
	\dot v^2 - \lambda v
	\label{}
\end{equation}
whose equation of motion is $\ddot v = - \frac{\lambda}{2}$, also obtainable from a Taylor expansion of \eqref{eq:ode}. With boundary conditions $v(0) = v(T) = 0$ the solution for $v(t)$ is:

\begin{equation}
	v(t) = - \frac{\lambda}{4} t^2 + \frac{\lambda}{4} T t\,.
	\label{}
\end{equation}
Applying the constraint $X = \int_0^T v dt = \frac{\lambda}{24} T^3$ the Lagrangian multiplier is determined to be $\lambda = 24 \, X / T^3$, so that the velocity profile $v(t; X,T)$ is

\begin{equation}
	\frac{v}{X/T} = -6 \left( \frac{t}{T} \right)^2 + 6 \frac{t}{T}\,,
	\label{}
\end{equation}
and the corresponding worldline is immediate by integration:

\begin{equation}
	\frac{x}{X} = - 2 \left( \frac{t}{T} \right)^3  + 3 \left( \frac{t}{T} \right)^2 + \operatorname{constant}\,
	\label{}
\end{equation}
matching with the result determined in \ADN{}.

While we regretably lack a closed-form expression for our solution, for a given distance and a desired proper time the minimal-discomfort trajectory can be determined through the constraints of integral equations \ref{eq:tintegral} and \ref{eq:xintegral}. These equations contain two unknowns, the maximum rapidity and the Lagrange multiplier. It is possible to solve these two equations simulataneously for the two unknowns, but difficult. Rather, the ratio of $X$ and $\mathcal{T}$ defines a characteristic ``average'' velocity (that may be superluminal), and dividing equation \ref{eq:tintegral} by equation \ref{eq:xintegral} 

\begin{multline}
	X/\mathcal{T} = \\  \int_0^{\rmax} \frac{\sinh s\,ds}{\sqrt{1- \dfrac{\sinh s}{\sinh \rmax}}} \,\Bigg/ 
		\int_0^{\rmax} \frac{ds}{\sqrt{1-\dfrac{\sinh s} {\sinh \rmax}  }} 
	\label{}
\end{multline}
fixes the relationship between $X/\mathcal{T}$ and $\rmax$, eliminating the dependence on $\lambda$, and can be more easily found numerically, being a one-dimensional relationship. Then, the desired proper time can be used to calculate the Lagrange multiplier using the known maximum rapidity. From this, the initial acceleration can be found iteratively.

Having derived the least uncomfortable trajectory for relativistic travel, we can calculate the rapidity as a function of proper time calculated by solving equation \ref{eq:ode} numerically using the Runge-Kutta algorithm, and use the integral equations \ref{eq:xintegral} and \ref{eq:tintegral} to constrain the Lagrange multiplier and the initial acceleration to the total distance and proper time.

\begin{figure}%
\includegraphics[width=1\columnwidth]{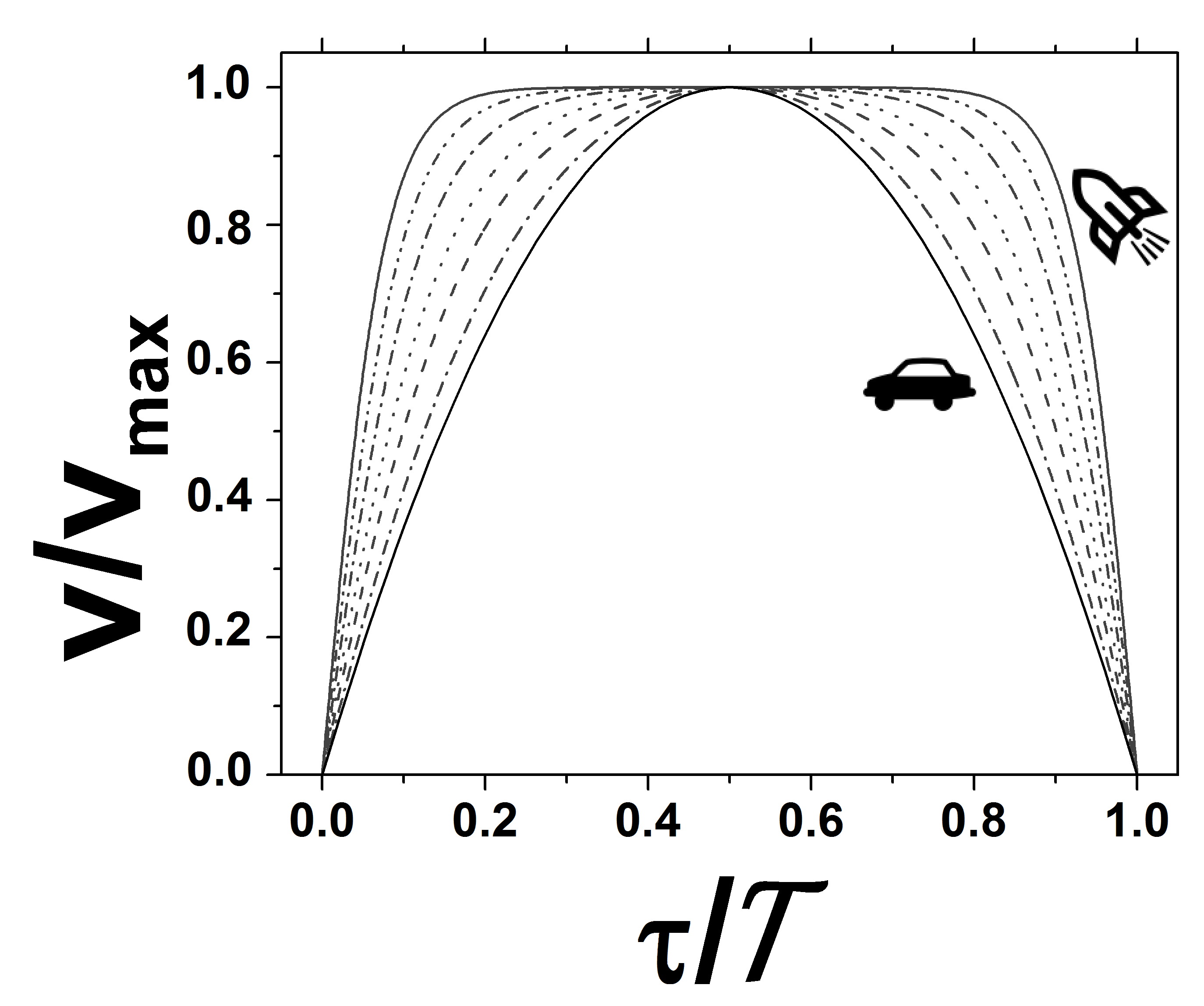}%
\caption{Velocity profiles of the least uncomfortable trips for varying maximum speeds from solutions to Equation \ref{eq:ode}. The inner solid curve denoted by the car is the classical solution Equation \ref{eq:classical}, with trips of increasing maximum speed characterized maximum Lorentz factor $\gamma_{max}$=1.4, 2.5, 4.1, 9.1, 25, and finally $\gamma_{max}$=100 denoted by the outer solid curve and the spaceship.}
\label{fig:vel}%
\end{figure}

We examine the trajectories in terms of the fractional velocity ($v/v_{max}$) as a function of relative proper time ($\tau/\mathcal{T}$) as seen in Fig. \ref{fig:vel}.  While the trajectory of the classical solution is universal, the relativistic solution is characterized by a single free parameter, the maximum speed (which, in effect, determines the degree of deviance from the classical solution), or equivalently the maximum rapidity or Lorentz factor. We classify solutions in terms of the maximum Lorentz factor, and our solutions reduce to the classical solution in the limit of $\gamma=1$. Because $v_{max}$ plateaus at $c$, the faster trajectories spend comparatively more of the trip near their maximum velocity. The behavior of the trajectories as the speed increases are similiar to solutions of the non-linear Schrodinger equation for repelling particles in a box, whose wavefunction changes from cosine-like to a more uniform distribution as the density increases \cite{carr}.

\begin{figure}%
\includegraphics[width=1\columnwidth]{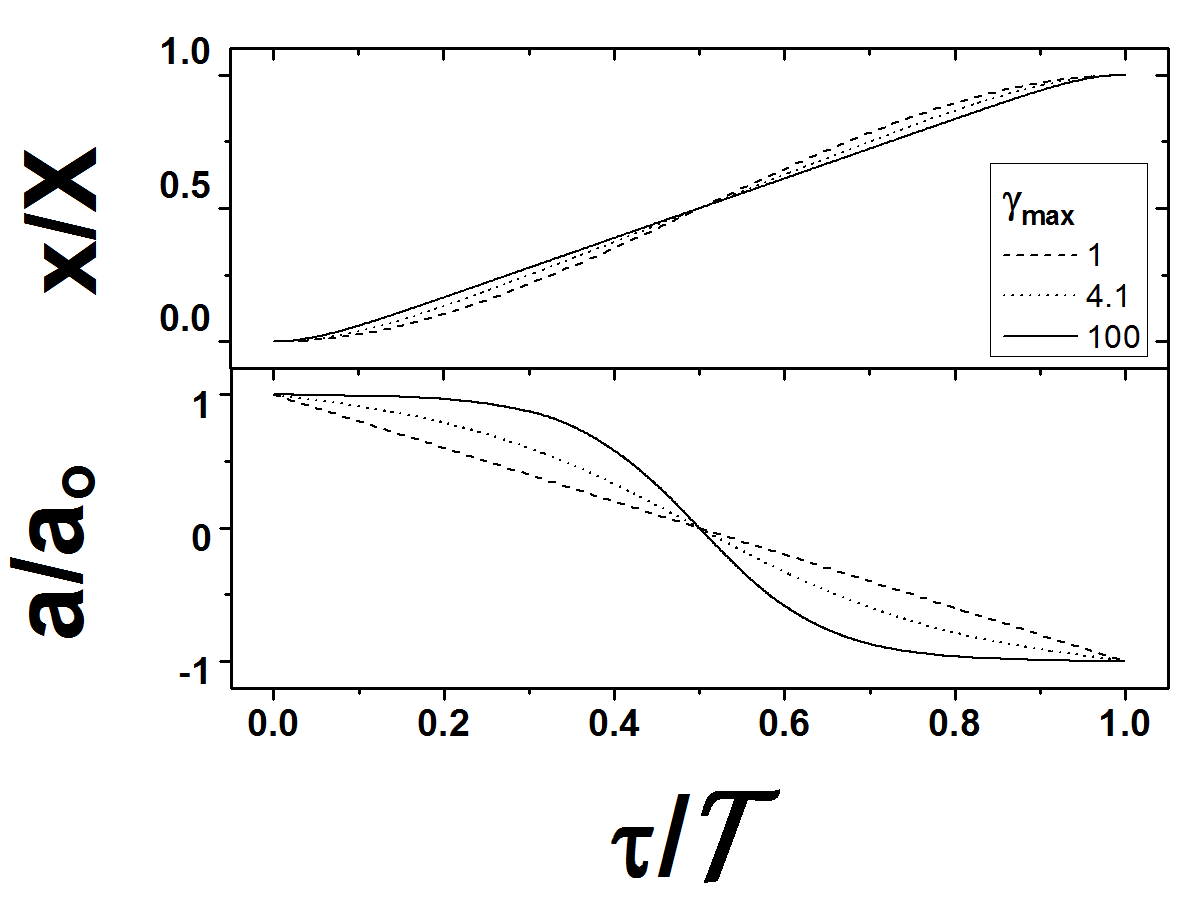}%
\caption{Top: Fractional distance as a function of fractional proper time for non-relativistic, slightly relativistic, and highly relativistic solutions. The trajectory is cubic at low velocities but with greater speed the trajectory approaches a line as $v\approx c$. Bottom: Proper acceleration relative to its initial value as a function of fractional time, for the same three cases. }%
\label{fig:xa}%
\end{figure}

The classical solution has a cubic time-evolution of position, implying that jerk is constant. This is not the case for the relativistic version; as the velocity approaches that of light it is nearly unchanging, and the trajectory approaches linearity (Fig. \ref{fig:xa}a). The coordinate acceleration thus approaches zero for most of the trip, however, the proper acceleration acquires a strong jerk as the maximum velocity increases (Fig. \ref{fig:xa}b).

\begin{figure}%
\includegraphics[width=1\columnwidth]{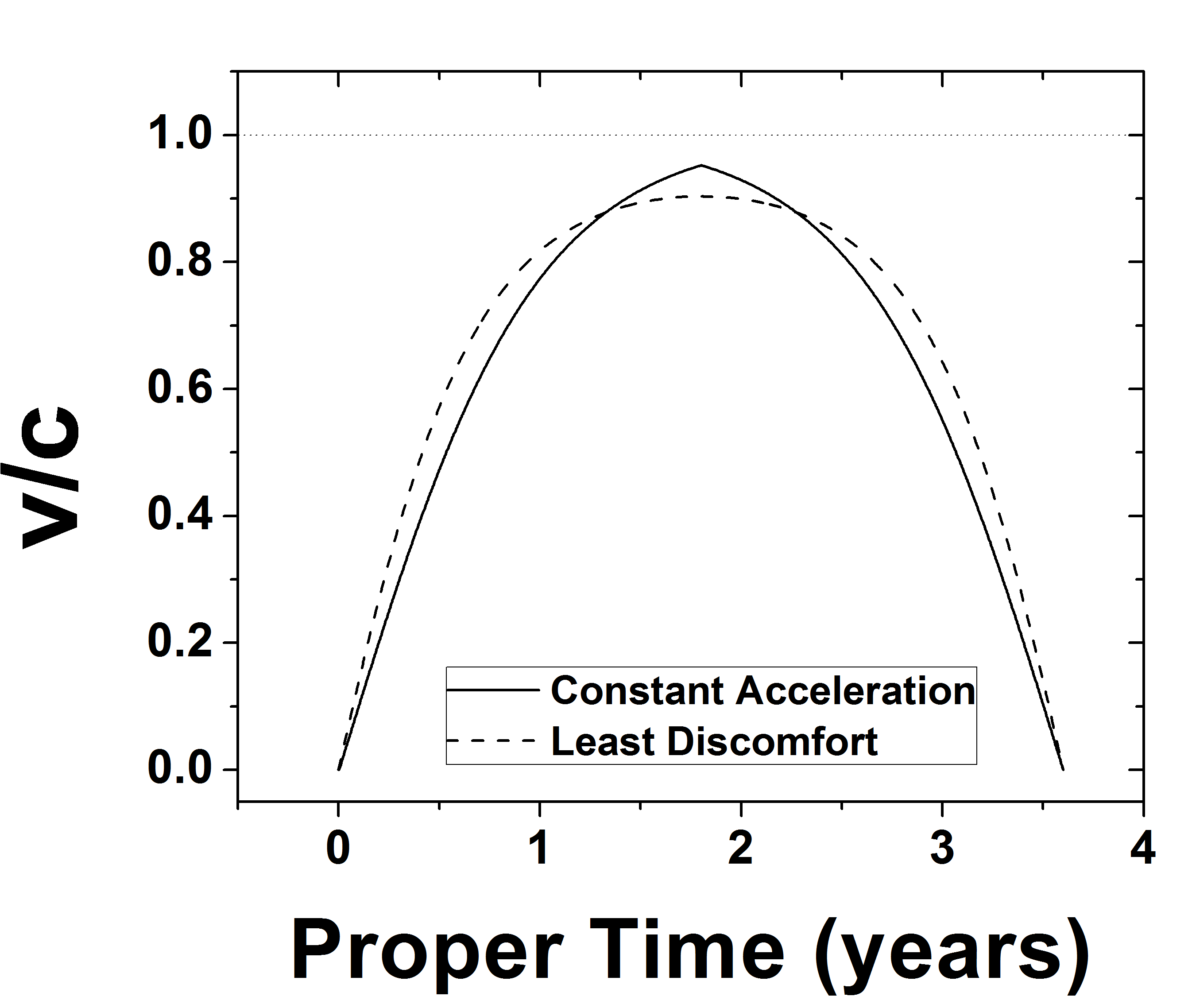}%
\caption{Two possible velocity profiles on a hypothetical trip to Alpha Centauri. The trip maintaining constant Earth-like proper acceleration reaches 95\% $c$ and has a kink discontinuity, while the minimal-discomfort scenario is smooth and reaches 90\% $c$. }%
\label{fig:comparison}%
\end{figure}

Analyses similar to ADN have been considered for the design of optimal train driving strategies \cite{pudney}. Trains and cars do not approach light speed and have a transverse source of terrestrial gravity, but on interstellar spaceflights for which this analysis may become relevant, a constant proper acceleration may be used as a source of artificial gravity. One could consider a trip to Alpha Centauri, for example, where +1 g of proper acceleration is applied for the first half of the trip, and -1 g is applied over the second half (with a brief period of nonzero jerk in the middle), coming to a rest at the destination while enjoying Earth-like gravity for (nearly) the entire trip. In a reference frame in which both the Earth and Alpha Centauri are close to being at rest, such a trip would take almost exactly six years through a pleasant coincidence, and a proper 3.6 years on-board, reaching 95 \% the speed of light. We can compare this trip to the minimal-discomfort trajectory with the proper time over the same distance (Fig. \ref{fig:comparison}). The minimal-discomfort trajectory only reaches 90 \% light speed, but spends more time closer to its maximum velocity, experiencing greater acceleration at the start and end of the trip. 

The constant-acceleration trip contains a kink discontinuity in the velocity corresponding to infinite jerk at the halfway point, as well as at the start and end of the journey. ADN \cite{anderson2016least} suggest an alternative comfort scheme, in which the squared jerk is minimized rather than the squared acceleration.  While this letter was motivated by the desire to see a relativistic generalization of the least-discomfort path, the most comfortable trajectory may be one that keeps gravity as close to g as possible while minimizing a higher-order kinematic derivative such as jerk. Considerations of jerk in special relativity have been explored previously \cite{sandin1990jerk, russo2009relativistic}, and minimizing the jerk for a relativistic journey is left as an exercise for the astute reader. We note that the minimal-discomfort path may be more suited to the transport of acceleration-sensitive equipment or self-replicating machines, rather than humans themselves.

We have extended the result of ADN \cite{anderson2016least} for the least uncomfortable linear trajectory to incorporate special relativity, and find a class of solutions that deviate from the classical solution as the maximum velocity approaches that of light. Although relativistic interstellar travel is likely many generations away, as physicists we believe it is not too early to consider the details of exotic-seeming transportation schemes \cite{klotz}. We hope that this work encourages students and researchers to consider the assumptions inherent in published results in physics, and to examine the implications of breaking those assumptions. 

\subsection*{Acknowledgements}
ARK is supported by an NSERC postdoctoral fellowship.

\bibliographystyle{unsrt}

\end{document}